\baselineskip=15pt
\magnification=\magstep1
\font\msbm=msbm10

\centerline{{\bf p-ADIC PATH INTEGRALS FOR QUADRATIC ACTIONS}}
\vskip1cm
\centerline{G. S. DJORDJEVI\'C$^*$ and B. DRAGOVICH}
\vskip.3cm
\centerline{{\it Institute of Physics, P.O. Box 57,}}
\centerline{{\it 11001 Belgrade, Yugoslavia}}
\centerline{{\it $^*$Department of Physics,University of Ni\v s,}}
\centerline{{\it P.O. Box 91, 18001 Ni\v s, Yugoslavia}}
\vskip1cm

\centerline{{\bf Abstract}}
\vskip.5cm

The Feynman path integral in p-adic quantum mechanics
is considered. The probability amplitude ${\cal
K}_p(x^{\prime\prime},t^{\prime\prime};x^\prime,t^\prime)$ for
one-dimensional systems with quadratic actions is calculated in an exact
form, which is the same as that in ordinary quantum mechanics.
\vskip1cm
\noindent
{\bf 1. Introduction}
\vskip.5cm

In the last decade years p-adic numbers have been successfully applied to
many parts of theoretical and mathematical physics (for a review, see,
e.g. Refs. 1-3).

As is known, numerical experimental results belong to the field of
rational numbers $\msbm\hbox{Q}$, and $\msbm\hbox{Q}$ is dense in
$\msbm\hbox{R}$ and in the field of p-adic numbers $\msbm\hbox{Q}_p$
($p$ is a prime number). $\msbm\hbox{R}$ and $\msbm\hbox{Q}_p$, for
every $p$, exhaust all possible number fields which can be obtained by
completing $\msbm\hbox{Q}$.

Any $x\in\msbm\hbox{Q}_p$ can be presented as an expansion
$$
x = p^\nu(x_0+x_1p+x_2p^2+\cdots)\ ,\quad \nu\in\msbm\hbox{Z}\ ,\eqno(1.1)
$$
where $x_i = 0,1,\cdots,p-1$. p-Adic norm of a term
$x_ip^{\nu+i}$ in (1.1) is $\mid x_ip^{\nu+i}\mid_p = p^{-(\nu+i)}$.
Since p-adic norm is the non-archimedean (ultrametric) one, {\it i.e.} $\mid
a+b\mid_p\leq\max \{\mid a\mid_p,\mid b\mid_p\}$, it means that the
canonical expansion (1.1) has $\mid x\mid_p = p^{-\nu}$. There is no
natural ordering on $\msbm\hbox{Q}_p$, but one can define a linear
order in the following way: $x<y$ if $\mid x\mid_p<\mid y\mid_p$, or
when $\mid x\mid_p = \mid y\mid_p$, there exists such index $m\geq0$
that digits satisfy $x_0 = y_0$, $x_1 = y_1,\cdots, x_{m-1} = y_{m-1}, x_m<y_m$.

There is an analysis over $\msbm\hbox{Q}_p$ connected with map
$\varphi:\msbm\hbox{Q}_p\to\msbm\hbox{Q}_p$, as well as another one
related to map $f:\msbm\hbox{Q}_p\to\msbm\hbox{C}$, where $\msbm\hbox{C}$
is the field of complex numbers. We use both of these analysis.
Derivatives$^4$ of $\varphi(x)$ are defined as in the real case, but with
respect to the p-adic norm. In the case of mapping
$\msbm\hbox{Q}_p\to\msbm\hbox{C}$ there is well-defined integral with
the Haar measure$^2$. In particular, we use the Gauss integral$^2$
$$
\int_{\msbm\hbox{Q}_{p}}\chi_p(\alpha x^2+\beta x)dx =
\lambda_p(\alpha)\mid
2\alpha\mid_p^{-1/2}\chi_p\bigg(-{\beta^2\over4\alpha}\bigg)\
,\quad\alpha\not=0\ ,\eqno(1.2)
$$
where $\chi_p(a) = \exp(2\pi i\{a\}_p)$ is an additive character and
$\{a\}_p$ is the fractional part of $a\in\msbm\hbox{Q}_p$.
$\lambda_p(x)$ is an arithmetic function, $\lambda_p(0) = 1$ and if
$x\in \msbm\hbox{Q}_p^*$ then
$$
\lambda_p(x) = \cases{1,&$\nu = 2k,\ \ \ \ \ \ \ p\not=2,$\cr
\big({x_0\over p}\big),&$\nu = 2k+1,\ \ p\equiv1(\hbox{mod 4}),$\cr
i\big({x_0\over p}\big),&$\nu = 2k+1,\ \ p\equiv3(\hbox{mod 4}),$\cr}\eqno(1.3)
$$ 
$$
\lambda_2(x) = \cases{{1\over\sqrt2}[1+(-1)^{x_{1}}i],&$\nu = 2k,$\cr
{1\over\sqrt2}(-1)^{x_{1}+x_{2}}[1+(-1)^{x_{1}}i],&$\nu = 2k+1,$\cr}\eqno(1.4)
$$
where $x$ is given by (1.1), $k\in\msbm\hbox{Z}$, and $\big({x_0\over
p}\big)$ is the Legendre symbol. The properties
$$
\eqalign{\lambda_p(0) = 1,\ \ \lambda_p(a^2x) = \lambda_p(x),\ \
&\lambda_p(x)\lambda_p(y) = \lambda_p(x+y)\lambda_p(x^{-1}+y^{-1}),\cr
&\lambda^*_p(x)\lambda_p(x) = 1\cr}\eqno(1.5)
$$
will be used.

One of the greatest achievements in the use of p-adic numbers in
physics is a formulation of p-adic quantum mechanics$^{5,6}$. The
elements of the corresponding Hilbert space ${\cal
L}_2(\msbm\hbox{Q}_p)$ are some complex-valued functions of a p-adic argument.
Quantization is performed by the Weyl procedure. Instead of the
Schr$\ddot {\rm o}$dinger equation, an eigenvalue problem and dynamical
evolution of a system are defined by means of an unitary
representation of the evolution operator $U_p(t)$ on ${\cal
L}_2(\msbm\hbox{Q}_p)$. Like ordinary quantum mechanics the kernel
${\cal K}_p(x^{\prime\prime},t^{\prime\prime};x^\prime,t^\prime)$ of
$U_p(t)$ is defined as
$$
U_p(t^{\prime\prime})\psi_p(x^{\prime\prime}) = \int_{\msbm\hbox{Q}_{p}}
{\cal
K}_p(x^{\prime\prime},t^{\prime\prime};x^\prime,t^\prime)\psi_p(x^\prime,t^\prime)dx^\prime\
.\eqno(1.6)
$$
As in the real case, ${\cal K}_p(x^{\prime\prime}
t^{\prime\prime};x^\prime,t^\prime)$ is called the quantum-mechanical propagator
or the probability amplitude for a quantum particle to go from a
space-time point $(x^\prime,t^\prime)$ to a space-time point
$(x^{\prime\prime},t^{\prime\prime})$. Anyhow, ${\cal
K}_p(x^{\prime\prime},t^{\prime\prime};x^\prime,t^\prime)$ is of
central importance in both ordinary and p-adic quantum mechanics.

For one-dimensional systems with quadratic Lagrangians it has been
assumed$^{7-10}$ that
$$
{\cal K}_p(x^{\prime\prime},t^{\prime\prime};x^\prime,t^\prime) = N_p(t^{\prime\prime},t^\prime)
\chi_p\bigg(-{1\over h}\bar
S(x^{\prime\prime},t^{\prime\prime};x^\prime,t^\prime)\bigg)\ ,\eqno(1.7)
$$
where $N_p(t^{\prime\prime},t^\prime)$ is a normalization factor, 
$\bar S(x^{\prime\prime},t^{\prime\prime};x^\prime,t^\prime)$ is a p-adic
classical action quadra-\break\noindent tic in $x^{\prime\prime}$ and $x^\prime$, and
$h$ is the Planck constant. It has been shown that (1.7) satisfies the
group property
$$
\int_{\msbm\hbox{Q}_{p}}{\cal K}_p(x^{\prime\prime},t^{\prime\prime};x,t)
{\cal K}_p(x,t;x^\prime,t^\prime)dx = {\cal
K}_p(x^{\prime\prime},t^{\prime\prime};x^\prime,t^\prime)\ ,\eqno(1.8)
$$
and $N_p(t^{\prime\prime},t^\prime)$ has been also calculated in an
explicit form for a harmonic oscillator$^{5,8}$, a free
particle$^{5,8}$, a particle in a constant field$^8$, a harmonic
oscillator with time-dependent frequency$^9$, a minisuperspace - the
de Sitter model of the universe$^{10}$, as well as for some other
minisuperspace cosmological models$^{11}$ with quadratic actions.

This Letter is devoted to p-adic generalization of the Feynman path
integral approach to ${\cal
K}(x^{\prime\prime},t^{\prime\prime};x^\prime,t^\prime)$ in ordinary
quantum mechanics. It will be derived in an exact
way for one-dimensional systems with quadratic classical actions, and
it will be also shown that such ${\cal
K}_p(x^{\prime\prime},t^{\prime\prime};x^\prime,t^\prime)$ has the same
form as that in ordinary quantum mechanics.

\vskip1cm
\noindent
{\bf 2. p-Adic Path Integrals}
\vskip.5cm

Since 1948, when Feynman published$^{12}$ his first paper on the path
integral, it has been a subject of permanent interest in theoretical
physics, and now it presents one of the best approaches to quantum
theory. On the recent progress, present status and some future
prospects of the path integral can be found Ref. 13.

In ordinary quantum mechanics, Feynman has postulated$^{12}$  
${\cal K}(x^{\prime\prime},t^{\prime\prime};x^\prime,t^\prime)$ to be
the path integral
$$
{\cal K}(x^{\prime\prime},t^{\prime\prime};x^\prime,t^\prime) = \int\exp\bigg(
{2\pi i\over h}S[q]\bigg){\cal D}q\ ,\eqno(2.1)
$$ 
where $S[q] = \int^{t^{\prime\prime}}_{t^{\prime}}L(q,\dot q,t)dt$ is
an action and $x^{\prime\prime} = q(t^{\prime\prime})$, $x^\prime =
q(t^\prime)$. The integral (2.1) symbolizes an intuitive understanding
that a quantum-mechanical particle may propagate from $x^\prime$ to
$x^{\prime\prime}$ using uncountably many paths which connect these two
points. Thus the Feynman path integral means a continual summation of
single amplitudes $\exp\bigg(
{2\pi i\over h}S[q]\bigg)$ over all paths $q(t)$ connecting $x^\prime$
and $x^{\prime\prime}$. In practical calculations it is the limit of an
ordinary multiple integral of $n-1$ variables $q_i = q(t_i)$ when
$n\to\infty$. For the classical action $\bar
S(x^{\prime\prime},t^{\prime\prime};x^\prime,t^\prime)$ which is polynomial
quadratic in $x^{\prime\prime}$ and $x^\prime$ it has been shown (see,
e.g. Ref. 14) that in ordinary quantum mechanics
$$
{\cal K}(x^{\prime\prime},t^{\prime\prime};x^\prime,t^\prime) = \bigg(
{i\over h}{\partial^2\bar S\over\partial x^{\prime\prime}\partial
x^\prime}\bigg)^{1/2}\exp\bigg({2\pi i\over h}\bar S
(x^{\prime\prime},t^{\prime\prime};x^\prime,t^\prime)\bigg)\ .\eqno(2.2)
$$

p-Adic generalization of (2.1) was suggested in Ref. 5 and one can
write it in the form
$$
{\cal K}_p(x^{\prime\prime},t^{\prime\prime};x^\prime,t^\prime) = \int
\chi_p\bigg(-{1\over h}S[q]\bigg){\cal D}q = \int\chi_p
\bigg(-{1\over h}\int^{t^{\prime\prime}}_{t^{\prime}}L(q,\dot
q,t)dt\bigg)\prod_tdq(t)\ ,\eqno(2.3)
$$
where $\chi_p(a)$ is p-adic additive character. In (2.3) we regard
$h\in\msbm\hbox{Q}$ and $q,t\in\msbm\hbox{Q}_p$. In fact, an
integral$^{15,16}$ $\int^{t^{\prime\prime}}_{t^{\prime}}L(q,\dot
q,t)dt$ is a difference of antiderivative of $L(q,\dot q,t)$ in points
$t^{\prime\prime}$ and $t^\prime$. However, $dq(t)$ is the Haar measure
and p-adic path integral is the limit of a multiple Haar integral.

The path integral (2.3) is elaborated, for the first time, for the
harmonic oscillator$^{16}$. In particular, it was shown that there
exists the limit
$$
\eqalign{{\cal K}_p(x^{\prime\prime},t^{\prime\prime};x^\prime,t^\prime) &=
\lim_{n\to\infty}{\cal
K}_p^{(n)}(x^{\prime\prime},t^{\prime\prime};x^\prime,t^\prime) =
\lim_{n\to\infty}N^{(n)}_p(t^{\prime\prime},t^\prime)\int_{\msbm\hbox{Q}_{p}}\cdots
\int_{\msbm\hbox{Q}_{p}}\cr
&\times\chi_p\bigg(-{1\over h}\sum^n_{i=1}\bar S(
q_i,t_i;q_{i-1},t_{i-1})\bigg)dq_1\cdots dq_{n-1}\ ,\cr}\eqno(2.4)
$$
where $N^{(n)}_p(t^{\prime\prime},t^\prime)$ is the corresponding
normalization factor for the harmonic oscillator. So obtained 
${\cal K}_p(x^{\prime\prime},t^{\prime\prime};x^\prime,t^\prime)$ is in
complete agreement with that earlier found in Refs. 5,8 by the method
described in Introduction.

In the similar way the path integral has been recently
calculated$^{17}$ for a particle in a constant external field. Also it
is done$^{18}$ for the linear oscillator with time-dependent frequency.
The obtained results confirm the form (1.7) and yield the corresponding
$N_p(t^{\prime\prime},t^\prime)$.

Thus one can conclude that p-adic path integrals calculated for
particular physical models give the results which resemble those known
in ordinary quantum mechanics. This is the reason to look for a general
expression in p-adic case which would be an analogue of (2.2) in the
real one. Note that in addition to (1.8), the path integral must also satisfy
$$
\int_{\msbm\hbox{Q}_{p}}{\cal K}_p^*(x^{\prime\prime},t^{\prime\prime};x^\prime,t^\prime)
{\cal K}_p(z,t^{\prime\prime};x^\prime,t^\prime)dx^\prime =
\delta_p(x^{\prime\prime}-z)\ ,\eqno(2.5)
$$
$$
{\cal K}_p(x^{\prime\prime},t^\prime;x^\prime,t^\prime) =
\lim_{t^{\prime\prime}\to t^{\prime}}{\cal K}_p(x^{\prime\prime},t^{\prime\prime};
x^\prime,t^\prime)= \delta_p(x^{\prime\prime}-x^\prime)\ ,\eqno(2.6)
$$
where $\delta_p(a-b)$ is the p-adic $\delta$-function$^2$.

\vskip1cm
\noindent
{\bf 3. General Solution for Quadratic Actions}
\vskip.5cm

We will derive now a general solution of the p-adic Feynman path
integral (2.3) for systems with Lagrangians $L(q,\dot q,t)$ , quadratic
in $q$ and $\dot q$, and classical actions $\bar
S(x^{\prime\prime},t^{\prime\prime};x^\prime,t^\prime)$, quadratic in
$x^{\prime\prime}$ and $x^\prime$.

Classical mechanics in p-adic case has the same analytic form as in the real one
(see Refs. 15,16). An analytic solution of the Euler-Lagrange equation $
{\partial L\over\partial q}-{d\over dt}{\partial L\over\partial\dot q}
= 0$ gives a p-adic classical path $\bar q(t)$. The corresponding
classical action is
$$
S[\bar q] = \bar S(x^{\prime\prime},t^{\prime\prime};x^\prime,t^\prime) = \int^{
t^{\prime\prime}}_{t^{\prime}}L(\bar q,\dot{\bar q},t)dt\ ,\quad
x^{\prime\prime} = \bar q(t^{\prime\prime})\ ,\ x^\prime = \bar
q(t^\prime)\ .\eqno(3.1)
$$

A possible quantum path $q = q(t)$ can be presented as $q(t) = \bar
q(t)+y(t)$ with conditions $y(t^\prime) = y(t^{\prime\prime}) = 0$.
Since $\delta S[\bar q] = 0$,  we can write for quadratic 
actions (Lagrangians) an expansion
$$
S[q] = S[\bar q+y] = S[\bar q]+{1\over2!}\delta^2S[\bar q] = S[\bar q]+{1\over2}
\int^{t^{\prime\prime}}_{t^{\prime}}\bigg(y{\partial\over\partial q}
+\dot y{\partial\over\partial\dot q}\bigg)^{(2)}L(q,\dot q,t)dt\ .\eqno(3.2)
$$
Replacing $S[q]$ by (3.2) and ${\cal D}q$ by ${\cal D}y$ in (2.3) we have
$$
{\cal K}_p(x^{\prime\prime},t^{\prime\prime};x^\prime,t^\prime) = 
\chi_p\bigg(-{1\over h}S[\bar q]\bigg)\int\chi_p\bigg(-{1\over2h}
\int^{t^{\prime\prime}}_{t^{\prime}}\big(y{\partial\over\partial
q}+\dot y{\partial\over\partial\dot q}\big)^{(2)}Ldt\bigg)\prod_tdy(t)\
.\eqno(3.3)
$$
Since the remained path integral in (3.3) does not depend on $x^\prime$
and $x^{\prime\prime}$ (see, e.g. Ref. 19) it follows that 
${\cal K}_p(x^{\prime\prime},t^{\prime\prime};x^\prime,t^\prime)$ has
the form (1.7).

Let us use the unitary condition (2.5) to get further information on
$N_p(t^{\prime\prime},t^\prime)$. After substitution (1.7) in (2.5) it becomes
$$
\mid N_p(t^{\prime\prime},t^\prime)\mid_\infty^2\int_{\msbm\hbox{Q}_{p}}\chi_p
[{1\over h}\bar S(x^{\prime\prime},t^{\prime\prime};x^\prime,t^\prime)-
{1\over h}\bar S(z,t^{\prime\prime};x^\prime,t^\prime)]dx^\prime = \delta_p
(x^{\prime\prime}-z)\ ,\eqno(3.4)
$$
where $\mid{}\mid_\infty$ denotes the usual absolute value. As a
consequence of quadratic dependence on $x^{\prime\prime},x^\prime$ and
$z$ one has
$$
\eqalign{&\bar
S(x^{\prime\prime},t^{\prime\prime};x^\prime,t^\prime)-\bar
S(z,t^{\prime\prime};x^\prime,t^\prime) =
(x^{\prime\prime}-z){\partial\over\partial z}\bar S(0,t^{\prime\prime};0,t^\prime)\cr
&+{1\over2!}[(x^{\prime\prime})^2-z^2]{\partial^2\over\partial z^2}\bar
S(0,t^{\prime\prime};0,t^\prime)+(x^{\prime\prime}-z)x^\prime{\partial^2
\over\partial
x^{\prime\prime}\partial x^\prime}\bar S
(0,t^{\prime\prime};0,t^\prime)\ .\cr}\eqno(3.5)
$$
Replacing (3.5) in (3.4) we obtain
$$
\eqalign{&\mid N_p(t^{\prime\prime},t^\prime)\mid_\infty^2\chi_p\bigg[{1\over
h}(x^{\prime\prime}-z){\partial\bar S\over\partial z}+{1\over2h}((x^{\prime\prime})^2-z^2)
{\partial^2\bar S\over\partial z^2}\bigg]\cr
&\times\delta_p\bigg(
(x^{\prime\prime}-z){1\over h}{\partial^2\bar S\over\partial
x^{\prime\prime}\partial x^\prime}\bigg) =
\delta_p(x^{\prime\prime}-z)\ .\cr}\eqno(3.6)
$$
Since $\chi_p$ and $\delta_p$-function depend on
$x^{\prime\prime}-z$ one can take $\chi_p(\cdots) = 1$ and
(3.6) leads to
$$
\mid N_p(t^{\prime\prime},t^\prime)\vert_\infty = \Big\arrowvert{1\over h}{\partial^2\over\partial
x^{\prime\prime}\partial x^\prime}\bar S(x^{\prime\prime},t^{\prime\prime};x^\prime,t^\prime)
\Big\arrowvert^{1/2}_p\ .\eqno(3.7)
$$

The general form of $N_p(t^{\prime\prime},t^\prime)$ in (1.7) is
$$
N_p(t^{\prime\prime},t^\prime) = \mid N_p(t^{\prime\prime},t^\prime)\vert_\infty
A_p(t^{\prime\prime},t^\prime)\ ,\eqno(3.8)
$$
where $A_p(t^{\prime\prime},t^\prime)$ is a complex-valued function of
$t^\prime, t^{\prime\prime}\in Q_p$ and
$\mid A_p(t^{\prime\prime},t^\prime)\mid_\infty = 1$. To determine
$A_p(t^{\prime\prime},t^\prime)$ we use the property (1.8) and the
Taylor expansions of quadratic actions, e.g.
$$
\eqalign{\bar S(x^{\prime\prime},t^{\prime\prime};x^\prime,t^\prime) &=
\bar S(0,t^{\prime\prime};0,t^\prime)+\bigg(x^{\prime\prime}{\partial\over\partial
x^{\prime\prime}}+x^\prime{\partial\over\partial x^\prime}
\bigg)\bar S(0,t^{\prime\prime};0,t^\prime)\cr
&+{1\over2!}
\bigg(x^{\prime\prime}{\partial\over\partial
x^{\prime\prime}}+x^\prime{\partial\over\partial x^\prime}
\bigg)^{(2)}\bar S(0,t^{\prime\prime};0,t^\prime)\ .\cr}\eqno(3.9)
$$

The integration in (1.8) is performed by applying the Gauss integral
(1.2). To satisfy (1.8), we have the following necessary and sufficient
conditions:
$$
A_p(t^{\prime\prime},t)A_p(t,t^\prime)\lambda_p(\alpha) =
A_p(t^{\prime\prime},t^\prime)\ ,\eqno(3.10)
$$
$$
\Big\arrowvert{1\over h}{\partial^2\bar S\over\partial
x^{\prime\prime}\partial x}\Big\arrowvert^{1/2}_p
\Big\arrowvert{1\over h}{\partial^2\bar S\over\partial
x\partial x^\prime}\Big\arrowvert^{1/2}_p\mid2\alpha\mid^{-1/2}_p = 
\Big\arrowvert{1\over h}{\partial^2\bar S\over\partial
x^{\prime\prime}\partial x^\prime}\Big\arrowvert^{1/2}_p\ ,\eqno(3.11)
$$
where
$$
\alpha = -{1\over 2h}{\partial^2\over\partial
x^2}\bar S(x^{\prime\prime},t^{\prime\prime};x,t)-
{1\over 2h}{\partial^2\over\partial
x^2}\bar S(x,t;x^\prime,t^\prime)\ .\eqno(3.12)
$$
Note that there is also another condition induced by an equality
between characters, but it does not affect (3.10) and (3.11), and
therefore its consideration will be omitted here. 

To satisfy (3.11) and to derive an expression for
$A_p(t^{\prime\prime},t^\prime)$ in (3.10) one has to adopt some p-adic
relations for derivatives of the classical action. It is natural to
start with equation (3.11) whose all ingredients are determined. It can
be rewritten in the form
$$
\eqalign{&\Bigg\arrowvert{-h\bigg({\partial^2\over\partial x^2}\bar S(x^{\prime\prime},t^{\prime\prime};x,t)\bigg)
+{\partial^2\over\partial x^2}\bar S(x,t;x^\prime,t^\prime)\over
{\partial^2\over\partial x^{\prime\prime}\partial x}\bar S(x^{\prime\prime},t^{\prime\prime};x,t)
{\partial^2\over\partial x\partial x^\prime}\bar S(x,t;x^\prime,t^\prime)}
\Bigg\arrowvert^{-1/2}_p\cr 
&= \Bigg\arrowvert h\bigg({\partial^2\over\partial x^{\prime\prime}\partial
x^\prime}\bar S(x^{\prime\prime},t^{\prime\prime};x^\prime,t^\prime)\bigg)^{-1}
\Bigg\arrowvert^{-1/2}_p\ .\cr}\eqno(3.13)
$$
The relations
$$
\eqalign{&{\partial^2\over\partial x^2}\bar S
(x^{\prime\prime},t^{\prime\prime};x,t)
+{\partial^2\over\partial x^2}\bar S(x,t;x^\prime,t^\prime)\cr
& = -u\bigg({\partial^2\over\partial x^{\prime\prime}\partial x}\bar S(x^{\prime\prime},t^{\prime\prime};x,t)
+{\partial^2\over\partial x\partial x^\prime}\bar S(x,t;x^\prime,t^\prime)
\bigg)\ ,\cr}\eqno(3.14)
$$
$$
\eqalign{\bigg({\partial^2\over\partial
x^{\prime\prime}\partial x}&\bar S(x^{\prime\prime},t^{\prime\prime};x,t)
\bigg)^{-1}+
\bigg({\partial^2\over\partial
x\partial x^\prime}\bar S(x,t;x^\prime,t^\prime)
\bigg)^{-1} \cr
&= v\bigg({\partial^2\over\partial
x^{\prime\prime}\partial x^\prime}\bar S(x^{\prime\prime},t^{\prime\prime};x^\prime,t^\prime)
\bigg)^{-1}\ ,\cr}\eqno(3.15)
$$
where $u$ and $v$ have for any particular $p$ expansions:
$$
u = 1+u_1p+u_2p^2+u_3p^3+\cdots\ ,\eqno(3.16)
$$
$$
v = 1+v_1p+v_2p^2+v_3p^3+\cdots\ ,\eqno(3.17)
$$
satisfy condition (3.11). If $u_1 = u_2 = v_1 = v_2 = 0$ for $p
= 2$ then by virtue of (1.3) and (1.4)
$$
\lambda_p(ux) = \lambda_p(vx) = \lambda_p(x)\eqno(3.18)
$$
for every $p$. Using condition (3.10), ralations (3.14), (3.15), and
properties (1.5), (3.18) of the $\lambda_p$ function it follows
$$
A_p(t^{\prime\prime},t^\prime) = \lambda_p\bigg(
-{1\over 2h}{\partial^2\over\partial
x^{\prime\prime}\partial x^\prime}\bar S(x^{\prime\prime},t^{\prime\prime};x^\prime,t^\prime)
\bigg)\ .\eqno(3.19)
$$

Finally we obtained the kernel of an evolution operator
$$
{\cal K}_p(x^{\prime\prime},t^{\prime\prime};x^\prime,t^\prime) = \lambda_p
\bigg(
-{1\over 2h}{\partial^2\bar S\over\partial
x^{\prime\prime}\partial x^\prime}
\bigg)
\Big\arrowvert{1\over h}{\partial^2\bar S\over\partial
x^{\prime\prime}\partial x^\prime}
\Big\arrowvert^{1/2}_p
\chi_p
\bigg(-{1\over h}\bar S(x^{\prime\prime},t^{\prime\prime};x^\prime,t^\prime)
\bigg)\eqno(3.20)
$$
as the solution of the p-adic Feynman path integral for quadratic actions.

One can easily verify (3.14) and (3.15) for a free particle, a particle
in a constant field and a harmonic oscillator.
\vskip1cm
\noindent
{\bf 4. Concluding Remarks}
\vskip.5cm

As has been mentioned earlier the Feynman path integral for quadratic
actions in the real case has an exact solution (2.2). It is worth
noting that (2.2) can be rewritten in the form
$$
{\cal K}_\infty(x^{\prime\prime},t^{\prime\prime};x^\prime,t^\prime) = \lambda_\infty
\bigg(
-{1\over 2h}{\partial^2\bar S\over\partial
x^{\prime\prime}\partial x^\prime}
\bigg)
\Big\arrowvert{1\over h}{\partial^2\bar S\over\partial
x^{\prime\prime}\partial x^\prime}
\Big\arrowvert^{1/2}_\infty
\chi_\infty
\bigg(-{1\over h}\bar S(x^{\prime\prime},t^{\prime\prime};x^\prime,t^\prime)
\bigg)\ ,\eqno(4.1)
$$
where index $\infty$ denotes the real case and
$$
\lambda_\infty(0) = 1\ ,\quad\lambda_\infty(a) = {1\over\sqrt2}(1-i \ \hbox{sign} \ a)\ ,
\quad a\in\msbm\hbox{R}^*
\eqno(4.2)
$$
is also an arithmetic function with the same properties (1.5). An
additive character on $\msbm\hbox{R}$ is defined as $\chi_\infty(x) = \exp
(-2\pi ix)$.

Let $v$ be an index which characterizes real and any of p-adic cases,
{\it i.e.} $v\in\{\infty,2,3,$ $5,\cdots\}$. Then the Feynman path
integral for a quantum-mechanical amplitude with qua-\break\noindent dratic classical
actions can be written in ordinary and p-adic quantum mechanics in the
same compact form
$$
{\cal K}_v(x^{\prime\prime},t^{\prime\prime};x^\prime,t^\prime) = \lambda_v
\bigg(
-{1\over 2h}{\partial^2\bar S\over\partial
x^{\prime\prime}\partial x^\prime}
\bigg)
\Big\arrowvert{1\over h}{\partial^2\bar S\over\partial
x^{\prime\prime}\partial x^\prime}
\Big\arrowvert^{1/2}_v
\chi_v
\bigg(-{1\over h}\bar S(x^{\prime\prime},t^{\prime\prime};x^\prime,t^\prime)
\bigg)\ .\eqno(4.3)
$$

Note that ${\cal
K}_v(x^{\prime\prime},t^{\prime\prime};x^\prime,t^\prime)$ is a
function of $t^{\prime\prime}-t^\prime$ if $L = L(q,\dot q)$.

The expression (4.3) exhibits a generic aspects of quantum particle
propagation in metric and ultrametric spaces. It once again underlines
the fundamental role of the Feynman path integral method in foundation
of quantum theory.

The obtained result (4.3) is also a solid basis for further elaboration
of Adelic quantum mechanics$^{20}$ (which unifies ordinary and p-adic
ones), for an approximate (semiclassical) computation of the p-adic path integrals in
the case of non-quadratic Lagrangians, and for a generalization to
multi-dimensional systems.
\vskip1cm
\noindent
{\bf References}

\item{1.} L. Brekke and P. G. O. Freund, {\it Phys. Rep.} {\bf 233}, 1 (1993).
\item{2.} V. S. Vladimirov, I. V. Volovich and E. I. Zelenov, {\it
p-Adic Analysis and Mathematical Physics} (World Scientific, Singapore,
1994).
\item{3.} A. Khrennikov, {\it p-Adic Valued Distributions in
Mathematical Physics} (Kluwer Acad. Publishers, Dordrecht, 1994).
\item{4.} W. H. Schikhof, {\it Ultrametric Calculus} (Cambridge Univ.
Press, 1984).
\item{5.} V. S. Vladimirov and I. V. Volovich, {\it Commun. Math.
Phys.} {\bf 123}, 659 (1989).
\item{6.} Ph. Ruelle, E. Thiran, D. Verstegen and J. Weyers, {\it J.
Math. Phys.} {\bf 30}, 2854 (1989).
\item{7.} P. G. O. Freund and M. Olson, {\it Nucl. Phys.} {\bf B297},
86 (1988).
\item{8.} C. Alacoque, P. Ruelle, E. Thiran, D. Verstegen and J.
Weyers, {\it Phys. Lett.} {\bf B211}, 59 (1988).
\item{9.} G. S. Djordjevi\'c and B. Dragovich, {\it
Facta Universitatis} {\bf 1} (3), 204 (1996).
\item{10.} B. Dragovich, {\t Adelic Wave Function of the Universe}, in
Proceedings of the Third A. Friedmann Intern. Seminar on Gravitation
and Cosmology, S. Petersburg, 1996.
\item{11.} B. Dragovich and Lj. Ne\v si\'c,  {\it Facta Universitatis}
{\bf 1} (3), 223 (1996).
\item{12.} R. P. Feynman, {\it Rev. Mod. Phys.} {\bf 20}, 367 (1948).
\item{13.} a special issue on functional integration, {\it J. Math. Phys.} {\bf 36} (5) (1995), .
\item{14.} C. Morette, {\it Phys. Rev.} {\bf 81}, 848 (1951).
\item{15.} B. Dragovich, P. H. Frampton and B. V. Urosevic, {\it Mod.
Phys. Lett.} {\bf A5}, 1521 (1990).
\item{16.} E. I. Zelenov, {\it J. Math. Phys.}, {\bf 32}, 147 (1991).
\item{17.} G. S. Djordjevi\'c and B. Dragovich, {\it On p-Adic
Functional Integration}, in Proceedings of the II Mathematical Conf. in
Pri\v stina, Pri\v stina (Yugoslavia), 1996.
\item{18.} G. S. Djordjevi\'c and B. Dragovich, {\it Some p-Adic and
Adelic Properties of the Harmonic Oscillator with Time-dependent
Frequency}, in preparation.
\item{19.} R. P. Feynman and A. Hibbs, {\it Quantum Mechanics and Path
Integrals} (McGraw Hill, New York, 1965).
\item{20.} B. Dragovich, {\it Int. J. Mod. Phys.} {\bf A10}, 2349 (1995).

\end